\documentclass[12pt]{article}

\usepackage{cite}

\usepackage{amssymb,latexsym}

\usepackage{epstopdf}

\usepackage{graphics,epsfig}

\usepackage{color}
\usepackage{xcolor}

\usepackage{amsmath,amsfonts}

\usepackage{mathrsfs}

\DeclareMathAlphabet{\mathpzc}{OT1}{pzc}{m}{it}

\let\a=\alpha \let\b=\beta \let\g=\gamma \let\d=\delta \let\e=\epsilon
\let\z=\zeta  \let\th=\theta  \let\k=\kappa
\let\l=\lambda \let\m=\mu \let\n=\nu \let\x=\xi \let\p=\pi 
\let\s=\sigma \let\t=\tau  \let\f=\phi  
 
        \let\Th=\Theta \let\L=\Lambda
\let\X=\Xi  \let\S=\Sigma  \let\Y=\Psi
 
\let\la=\label  
  
\def\nn{\nonumber} \def\bd{\begin{document}} \def\ed{\end{document}}
\def\ds{\documentstyle} \let\fr=\frac \let\bl=\bigl \let\br=\bigr
\let\Br=\Bigr \let\Bl=\Bigl
\let\bm=\bibitem
\let\na=\nabla
\def\tU{{\widetilde U}}
\let\pa=\partial \let\ov=\overline
\def\ie{{\it i.e.\ }}
\newcommand{\be}{\begin{equation}}
\newcommand{\ee}{\end{equation}}
\def\ba{\begin{array}}
\def\ea{\end{array}}
\def\ft#1#2{{\textstyle{{\scriptstyle #1}\over {\scriptstyle #2}}}}
\def\fft#1#2{{#1 \over #2}}
\def\F#1#2{{ F_{#1}^{(#2)} }}
\def\cF#1#2{{ {\cal F}_{#1}^{(#2)} }}

\def\R{{\bf R}}
\def\sst#1{{\scriptscriptstyle #1}}
\def\oneone{\rlap 1\mkern4mu{\rm l}}
\def\e7{E_{7(+7)}}
\def\td{\tilde}
\def\wtd{\widetilde}
\def\im{{\rm i}}
\def\bog{Bogomol'nyi\ }
\newcommand{\ho}[1]{$\, ^{#1}$}
\newcommand{\hoch}[1]{$\, ^{#1}$}
\newcommand{\bea}{\begin{eqnarray}}
\newcommand{\eea}{\end{eqnarray}}
\newcommand{\ra}{\rightarrow}
\newcommand{\lra}{\longrightarrow}
\newcommand{\Lra}{\Leftrightarrow}
\newcommand{\ap}{\alpha^\prime}
\newcommand{\bp}{\tilde \beta^\prime}
\newcommand{\cB}{{\cal B}}
\newcommand{\cO}{{\cal O}}
\newcommand{\vecx}{\vec{x}}
\newcommand{\vecy}{\vec{y}}
\newcommand{\vecp}{\vec{p}}
\newcommand{\vecq}{\vec{q}}
\newcommand{\tr}{{\rm tr} }
\newcommand{\Tr}{{\rm Tr} }
\newcommand{\NP}{Nucl. Phys. }

\newcommand{\cL}{{\cal L}}
\newcommand{\cA}{{\cal A}}
\newcommand{\cT}{{\cal T}}
\newcommand{\cR}{{\cal R}}
\newcommand{\cD}{{\cal D}}
\newcommand{\cH}{{\cal H}}

\def\Cb{\bar{C}}

\def\sst#1{{\scriptscriptstyle #1}}
\def\0{{\sst{(0)}}}
\def\1{{\sst{(1)}}}
\def\2{{\sst{(2)}}}
\def\3{{\sst{(3)}}}
\def\4{{\sst{(4)}}}
\def\5{{\sst{(5)}}}
\def\6{{\sst{(6)}}}
\def\7{{\sst{(7)}}}
\def\8{{\sst{(8)}}}
\def\9{{\sst{(9)}}}
\def\p{{\sst{(p)}}}
\def\q{{\sst{(q)}}}
\def\ve{\varepsilon}
\def\vf{\varphi}
\def\F{\Phi}
\def\wg{\wedge}

\def\thb{\bar{\theta}}
\def\Thb{\bar{\Theta}}
\def\barp{\bar{p}}
\def\barq{\bar{q}}
\def\barc{\bar{c}}
\def\bard{\bar{d}}
\def\e{\epsilon}

\def \bi{\bibitem}
\def \la {\label}

\def \l {\lambda}
\def\foot{\footnote}
\def \tl  {{\tilde \l}}
\def \sql {{\sqrt \l}}
\def \adss {$AdS_5 \times S^5$\ }
\newcommand{\rf}[1]{(\ref{#1})}
\def \ov {\over}

\def\th{\theta}
\def\Th{\Theta}
\def\vth{\vartheta}
\def\btheta{{\bar\theta}}
\def\ttheta{{{\tilde\theta}}}
\def\bttheta{{{\bar\ttheta}}}
\def\vth{\vartheta}

\def\ra{\rightarrow}
\def\N{\nabla}
\def\F{{\cal F}}
\def\uM{\underline{M}}
\def\uA{\underline{A}}
\def\uN{\underline{N}}
\def\uP{\underline{P}}
\def\ua{\underline{a}}
\def\ub{\underline{b}}
\def\uc{\underline{c}}
\def\ud{\underline{d}}
\def\ue{\underline{e}}
\def\uf{\underline{f}}
\def\ui{\underline{i}}
\def\uj{\underline{j}}
\def\uk{\underline{k}}
\def\ul{\underline{l}}
\def\ual{\underline{\alpha}}
\def\ube{\underline{\beta}}
\def\um{\underline{m}}
\def\un{\underline{n}}
\def\up{\underline{p}}
\def\uq{\underline{q}}
\def\ur{\underline{r}}
\def\us{\underline{s}}
\def\umu{\underline{\mu}}
\def\unu{\underline{\nu}}
\def\ula{\underline{\l}}
\def\uka{\underline{\k}}
\def\usi{\underline{\s}}
\def\urh{\underline{\r}}
\def\cc{\circ}
\def\eqv{\equiv}

\def\ni{\noindent}

\def\Ep{E^{{}^{(+)}}}
\def\Em{E^{{}^{(-)}}}

\def\Mp{M^{{}^{(+)}}}
\def\Mm{M^{{}^{(-)}}}

\def \ha{{1\ov 2}}

\def\r{\rho}

\def\Y{{\rm Y}}
\def\X{{\rm X}}
\def\tY{\tilde{\rm Y}}
\def\tX{\tilde{\rm X}}
\def\dY{\dot{\rm Y}}
\def\dX{\dot{\rm X}}

\def \J {\mathcal{J}}
\def \del {\partial}

\def\dF{\dot{F}}
\def\dG{\dot{G}}
\def\df{\dot{f}}
\def \E {{\cal E}}
\def \S {{\cal S}}
\def \J {{\cal J}}

\def\ms{\mathcal{S}}
\def\mj{\mathcal{J}}
\def\soj{\fr{\ms}{\mj}}
\def \R {{\bf R}}
\def \om {\omega}
\def \bE {\bar E}
\def \x {{\cal X}}

\def \bi{\bibitem}
\def \la {\label}

\def \l {\lambda}
\def\foot{\footnote}
\def \tl  {{\tilde \l}}
\def \sql {{\sqrt \l}}
\def \adss {$AdS_5 \times S^5$\ }
\def \ov {\over}

\def \varpi {{\rm w}}

\def\thb{\bar{\theta}}
\def\Thb{\bar{\Theta}}
\def\mb{\bar{\m}}
\def\ab{\bar{\a}}
\def\zb{\bar{z}}
\def\psib{\bar{\psi}}
\def\barp{\bar{p}}
\def\barq{\bar{q}}
\def\barc{\bar{c}}
\def\bard{\bar{d}}
\def\e{\epsilon}
\def\wb{\bar{w}}
\def\lb{\bar{\l}}
\def\Jb{\bar{J}}
\def\Nb{\bar{N}}
\def\Zb{\bar{Z}}
\def\pab{\bar{\pa}}

\def\At{\tilde{A}}
\def\Bt{\tilde{B}}
\def\Ct{\tilde{C}}
\def\Dt{\tilde{D}}
\def\Et{\tilde{E}}
\def\Ft{\tilde{F}}
\def\Gt{\tilde{G}}
\def\Ht{\tilde{H}}
\def\Kt{\tilde{K}}
\def\Mt{\tilde{M}}
\def\Nt{\tilde{N}}
\def\Rt{\tilde{R}}
\def\at{\tilde{a}}
\def\bt{\tilde{b}}
\def\ct{\tilde{c}}
\def\dt{\tilde{d}}
\def\et{\tilde{e}}
\def\ft{\tilde{f}}
\def \ztt{\tilde{\z}}
\def \zetat{\tilde{\zeta}}
\def\htil{\tilde{h}}
\def\gt{\tilde{g}}
\def\nt{\tilde{n}}
\def\mut{\tilde{\mu}}
\def\nut{\tilde{\nu}}
\def\pht{\tilde{\f}}
\def\Phit{\tilde{\Phi}}
\def\vft{\tilde{\vf}}

\def\rht{\tilde{\rho}}

\def\asth{\hat{*}}
\def\phh{\hat{\phi}}

\def\bA{{\bf A}}

\def\ola{\overleftarrow}
\def\ora{\overrightarrow}
\def\alt{\tilde{\a}}

\def\eh{\hat{e}}
\def\eph{\hat{\e}}
\def\ph{\hat{p}}
\def\alh{\hat{\a}}
\def\beh{\hat{\b}}
\def\gah{\hat{\g}}
\def\Fh{\hat{F}}
\def\muh{\hat{\m}}
\def\nuh{\hat{\n}}
\def\thh{\hat{\th}}
\def\rhh{\hat{\r}}
\def\dh{\hat{d}}
\def\ih{\hat{i}}
\def\jh{\hat{j}}
\def\hh{\hat{h}}
\def\nh{\hat{n}}
\def\gh{\hat{g}}
\def\kh{\hat{k}}
\def\deh{\hat{\d}}
\def\wh{\hat{w}}
\def\lah{\hat{\l}}
\def\Ah{\hat{A}}
\def\Gh{\hat{G}}
\def\Kh{\hat{K}}
\def\Nh{\hat{N}}
\def\Rh{\hat{R}}
\def\Ch{\hat{C}}
\def\Omh{\hat{\Omega}}

\def\xh{\hat{x}}

\def\ps{\rlap{\, /}\;\,p }
\def\ks{\rlap{\, /}\;\,k }

\def\gym{g_{YM}}

\def\adot{\dot{a}}
\def\bdot{\dot{b}}
\def\bpa{\bar{\pa}}

\def\pr{\prime}
\def\ssk{\medskip}
\def\clb{\color{blue}}
\def\clr{\color{red}}
\def\clg{\color{green}}
\def\clp{\color{purple}}
\def\clc{\color{cyan}}
\def\clm{\color{magenta}}
\def\cly{\color{yellow}}

\def\bfA{{\bf A}}
\def\bfB{{\bf B}}
\def\bfK{{\bf K}}
\def\bfU{{\bf U}}
\def\bfX{{\bf X}}
\def\bfY{{\bf Y}}
\def\bfZ{{\bf Z}}
\def\bfg{{\bf g}}
\def\bfn{{\bf n}}

\def\bsk{\bigskip}
\def\ssk{\medskip}

\def\Ec{{\cal E}}

\begin{document}

\overfullrule=0pt
\parskip=2pt
\parindent=12pt
\headheight=0in \headsep=0in \topmargin=0in
\oddsidemargin=0in

\vspace{ -3cm}
\thispagestyle{empty}

 \vspace{0.1cm}

\setcounter{equation}{0}
\setcounter{footnote}{0}
\setcounter{section}{0}

\begin{center}

{\Large\bf Black hole evolution in quantum-gravitational framework}

\vskip 0.8cm

\vspace{0.5cm}

 I. Y. Park
\\

\vspace{0.3cm}

\vspace{0.3cm}
{\it Department of Applied Mathematics,
Philander Smith College 
                               \\
Little Rock, AR 72223, USA \\
inyongpark05@gmail.com
}

 \vspace{.5cm}

\end{center}

 \vspace{0.1cm}

\begin{abstract}

We found black hole evolution on a quantum-gravitational scattering framework with an aim to tackle the black hole information paradox. With this setup, various pieces of the system information are explicit from the start and unitary evolution is manifest throughout. The scattering amplitudes factorize into the perturbative part and nonperturbative part. The nonperturbative part is dominated by an instanton-type contribution, i.e., a black hole analogue of the Coleman-De Luccia's bounce solution, and we propose that the Hawking radiation be identified with the particles generated by the vacuum decay. Our results indicate that the black hole degrees of freedom are entangled not only with the Hawking modes but also with the pre-Hawking modes. The Wald's entropy charge measures their entanglement. The full quantum-gravitational entropy is defined as the vev of the Wald entropy charge. With this definition a {\em shifted} Page-like curve is generically generated and its quantum extension is readily defined.

\end{abstract}
\newpage





\section{Introduction}

Lack of quantizing gravity in a renormalizable manner has long been an obstacle to progress that would have otherwise been  possible. Various reviews of approaches to quantization can be found in \cite{DeWitt:1975ys,'tHooft:1974bx,Stelle:1976gc,Antoniadis:1986tu,Weinberg3,Reuter:1996cp,Odintsov:1990qq,Thiemann:2007zz,Ambjorn:2012jv,Donoghue:2015hwa,Buchbinder,Birrell,Esposito,Parker}. Due to the series of recent developments in gravity quantization in which renormalizability of the physical states has been established \cite{Park:2016zgt,Park:2019lkh,Park:2018vci}, however, one is no longer shackled by the offshell non-renormalizability.
The Firewall argument \cite{Almheiri:2012rt}\cite{Braunstein:2009my} is a recent reminder that all is not well in our previous understanding of physics involving gravitational phenomena, in particular, black hole physics. (Reviews on black hole physics can be found in \cite{Frolov,Mukhanov,Maggiore}.) 
Arguably, the most significant contribution of the Firewall is that it has shaken several widely adopted views in the field, leading to various further developments. One of the views that the Firewall argument has put under close scrutiny is the conventional interpretation of Equivalence Principle as implying a smooth and structureless horizon. The purpose of the present work is twofold: the first is to collect various ideas in the literature related to black hole information  \cite{Hawking:1974sw} (reviews can be found in \cite{Page:1993up,Strominger:1994tn,Susskind,Mathur:2009hf}) and Firewall, and put them in a quantum-gravitational perspective. 
The second is to present a quantum-gravitational framework of black hole evolution, i.e., formation and evaporation, with an aim to tackle the black hole information (BHI) paradox among other things. In particular, the ideas put forth in \cite{Park:2013rm,Park:2017wiw} on the relevance of the pre-Hawking radiation and vacuum decay to BHI are made precise.

The quantum gravitational setup to tackle BHI that we employ is along the line of the scattering amplitude framework \cite{tHooft:1996rdg}, and in that sense it is not new. However, we will pay special attention to the pre-Hawking modes and boundary dynamics as well as the role of a bounce worm hole solution. In our analysis of the black hole evolution, all of the types of the system information are manifestly represented from the outset: the pieces of the information carried by various components appearing in the scattering amplitudes, such as the nonperturbative contributions, are explicitly (at least, conceptually) exhibited. The system information is classified into three categories: the perturbative information associated with the Fock oscillators, the information associated with the instanton-type non-perturbative contribution, and the information that can be extracted through a black hole perturbation analysis. All of these pieces of the information are preserved, and unitary evolution is maintained throughout.

The focus of the quantitative analysis of the present work is the construction of the bounce solution, computation of its entropy, and its role in the information paradox. These are central to the resolution of the BHI paradox. (For the other two categories, we discuss their status and tasks that need to be carried out for a complete picture.) The non-perturbative contribution to the path integral is dominated by a solution that is nothing but a black hole analogue of the Coleman-De Luccia's bounce solution \cite{Coleman:1980aw}\cite{Weinberg}. Although the connection to the Coleman-De Luccia's bounce solution is newly proposed in the present work (as far as we can see), the same name, ``bounce," has been used for the ``Lorentzian bounce solutions" in the gravitational context \cite{Ambrus:2005nm}\cite{Haggard:2014rza}.

There are several critical ingredients that have led to the outcome of the present work. Firstly, it is the recognition of the existence and role of the ``pre-Hawking" radiation \cite{Park:2013rm}. The pre-Hawking radiation, noted by Hawking himself \cite{Hawking:1974sw}, is an important part of the system components, and will be one of the main players in the entanglement of the system components. In addition to the information perturbatively stored by the Fock oscillators acting on the states, a substantial part of the information is stored in the entanglement between the components of the system, and to properly account for the entanglement, all of the system components must be counted. Secondly, it is the identification of the Hawking radiation: we propose that the particles generated by the vacuum decay be identified with the Hawking radiation. Particle generation by a time-dependent geometry is a well-known phenomenon \cite{Parker}. In the present context it is the bounce solution that generates the particles that in turn are identified with the Hawking particles. Although being overly simple and thus having some undesirable features, it captures the essence of the mechanism of information release. Thirdly, it is the relevance of the boundary degrees of freedom and their dynamics. (An earlier related discussion on the boundary dynamics in the context of BHI can be found in \cite{Marolf:2008mf}.) It has been competently demonstrated by a variant of Kaluza-Klein reduction that the boundary has nontrivial dynamics \cite{Park:2018xtt}, and the Hilbert space of the bulk theory must be enlarged accordingly \cite{Park:2016fxc} (see \cite{Freidel:2016bxd} for an earlier related observation), which in turn is crucial for gravity quantization and surrounding matters including BHI. These ingredients naturally become relevant as we proceed with the quantum gravitational setup presented in the main body.

The boundary configuration plays a critical role in choosing the in- and out- vacuua. To study the black hole forming process and its information perspective, the in-state is chosen to be a white hole and the out-state is built on a collapsing shell or shock-wave (not the other way around; see the comments in section 3). As we will detail, the states of the boundary theory dictates the bulk boundary conditions. The non-perturbative contribution to the path integral is dominated by the bounce solution. It represents a shell that collapses from an initial boundary condition associated with the white hole and subsequently expands; it bridges the vacuum transition {from the white hole state to the black hole state.} {Our toy model presented in the main body serves the purpose of illustrating the key ideas, with minimal technical complications.\footnote{{It would be ideal to construct a more realistic bounce solution of, say, an Einstein-scalar system with a scalar potential that realizes such mediation of the vacuum transition. Without such a solution available, we consider transitions among the configurations in the general solution space. More on this below.}} }

Several remarks are in order for the bounce solution and its physics. One thing worth noting is that in the present description of the Hawking radiation it is not necessary to construct a bulk solution that explicitly realizes a shrinking and evaporating black hole. This part of the physics is accounted for by the bounce wormhole solution as a non-perturbative contribution to the path integral. (See the related comments in section 3.3.) More specifically, it is the white whole state that accounts for the evaporated state: when the black hole evaporates, the mass is not lost, in the case of a Dirichlet boundary condition - but converted into the energy of the white hole state. In other words, it is the novelty of vacuum decay physics that allows one to bypass explicit construction of a shrinking bulk solution. Another interesting feature of our analysis is that Wald entropy computed by using the bounce solution reveals a ``shifted behavior", compared to the Page curve.\footnote{I greatly benefited from discussing with S. Braunstein on this.} For instance, the Wald entropy takes a maximum value when the area of the bounce solution is at maximum whereas the Page curve starts there (namely, takes zero). The fact that there is such a shift is natural, given that the Wald's entropy should count the entanglement from the pre-Hawking radiation as well. We expect that this feature will remain valid even for more realistic bounce solutions: the feature should be taken as a prediction of the present vantage point.

\vspace{.3in}

The rest of the paper is organized as follows. In section 2 we start by surveying some of the salient issues regarding black hole evolution and information. We try to connect various ideas on BHI in the literature, especially the ones that led to a non-smooth horizon. In section 3, we first present the setup of the scattering of particles around a black hole. The nonperturbative decaying process requires a careful examination. For that, we consider a simple bounce solution to avoid technical intricacies. As it stands our solution is Lorentizian; Euclideanization is briefly discussed. Despite the fact that it has certain undesirable features, it captures at least some of the essential aspects of the entanglement through the Wald's entropy. We discuss the issues surrounding the solution, and comment on a more realistic bounce solution.
In section 4, we ponder different types of black hole hair. The hair associated with the entanglement among {\em all} of the system components is crucial for the information problem. It was anticipated in \cite{Park:2017wiw} that a vacuum decay will play an important role in BHI. We make this idea concrete. The Wald's Noether charge measures the entanglement between the black hole degrees of freedom and the rest of the system \cite{Faulkner:2013ica}\cite{Oh:2017pkr}. (See also an earlier related work \cite{Brustein:2007jj}.) The bounce solution will be the main contributor to the Wald entropy, and a ``Page-like" curve is generically generated. We note, and explain the reason, that the Wald entropy becomes time-dependent, despite the fact that it is a Noether charge. The Wald entropy computed has an important difference, compared to the Page curve \cite{Page:1993wv}: it is shifted. For quantum-field-theoretic extension of the Wald's entropy we consider its vev; the result will correspond to the quantum-corrected entropy. We also draw a connection with recent works in which the Page curve \cite{Page:1993wv} was produced. Section 5 has a summary and discussions of possible implications of our results. We end with future directions. Appendix A contains some details regarding the bounce solution \rf{fsol} in section 3 and its generalization.

\section{Salient issue surrounding BHI}

In this section we will have a bird's eye view of some of the salient issues regarding black hole evolution and information. We relate to some of the ideas on a non-smooth horizon and BHI in the literature. The vacuum decaying process requires a more extensive treatment and will be taken up in the next section.

Although the Hawking's original work employs a collapsing shell, a simpler setup for demonstrating evaporation of a black hole is through an eternal black hole. Let us remind, by taking an eternal Schwarzschild black hole, the backbone of the BHI paradox in the semi-classical description. Afterwards we will turn to a time-dependent case.
As well known, there are several well-established coordinate systems, such as the Schwarzschild and Kruskal-Szekeres (Kruskal, for short) coordinates, adapted to observers in their states of the motion. Let us consider a free massless scalar field in the Schwarzschild and Kruskal coordinates, respectively. After semi-classical quantization of the scalar field, what leads to the Hawking radiation is the fact that the vacuua defined by the Schwarzschild observer and Kruskal observer are inequivalent \cite{Unruh:1976db}\cite{Birrell}. Due to this, the Kruskal vacuum will appear radiating to the Schwarzschild observer. The radiation is thermal; the black hole will continue radiating until it disappears, leaving the thermal radiation behind. At face value the picture implies that an initial pure state has evolved into a mixed state, thus violating unitarity; the information seems lost.

There are several spots needing improvement in the account above. For instance, given that the evolution of a pure state to a mixed state characterizes the loss of the information, is all of the system information lost? Where did the system information associated with the constituents of the infalling body (e.g., a collapsing shell) go? Even prior to these questions, what is the system information in the first place, since it must be more than the information carried by the entanglement among the constituents (although that part is central to the BHI paradox)? These questions suggest that a precise definition of the system information is desirable, and this is one of the things built in the fully quantum-gravitational setup. As we will see, it is the information associated with the entanglement that is hardest to see preserved whereas preservation of the rest of the  information is not as hard. Another less obvious place for improvement is the manner in which the semi-classical analysis treats the boundary dynamics: it simply misses the dynamics entirely. A series of recent works \cite{Hatefi:2012bp,Park:2016fxc,Park:2017wiw,Park:2018xtt} show that the boundary dynamics are critical for the information problem as well as gravity quantization. As we will see, the boundary dynamics\footnote{The boundary degrees of freedom seem to correspond to the photons mentioned in \cite{Unruh:2017uaw} as associated with ``innocuous information loss."} play a crucial role in resolving the paradox.

Various works in the past, including Hawking's original work \cite{Hawking:1974sw}, show that certain pieces of information escape at various stages of black hole evolution. In \cite{Hawking:1974sw} it was noted that prior to the Hawking radiation, radiation in super-radiant modes comes out.\footnote{Not being aware of this, a similar idea of pre-Hawking radiation was put forth in \cite{Park:2013rm}.} (These types of the pre-Hawking radiations will be important later in our picture of the proposed resolution of the paradox.) The Hawking radiation emerges only at the very late stage of the collapse \cite{Jacobson:1991gr}. To identify various pieces of information involved in the black hole information problem, it is necessary to carry out the analysis at various levels: the Hawking-type analysis of the kinetic terms, quantum-field-theoretic analysis including the interactions, and analysis of nonperturbative contributions. In sections 3 and 4 we will systematically place these in the overarching framework of the scattering theory.

As will be discussed in section 3, the escape of some pieces of the information can be easily recognized. This, however, is not the case for the Hawking radiation. Although it has long been expected that the Hawking radiation be purified by some other components of the system, the detailed mechanism has not been clearly revealed. The puzzle culminates with the Firewall argument: when crossing the horizon of an old black hole, an infalling observer experiences a trans-Planckian radiation.

Were it not for the Equivalence Principle, the Firewall would perhaps not be entirely surprising since the Schwarzschild vacuum would appear radiating to an infalling observer and the energy of the radiation should become more and more blue-shifted as the observer falls deeper on. Since the entanglement is expected to play an important role and constitutes a substantial part of the system information, one may investigate deeper into its involvement at the microscopic level. It is possible to understand this through a more systematic setup by employing a tripartite system.    
This was what was done in the Firewall argument. (As a matter of fact, there exist pre-Firewall indications of a non-smooth horizon in the literature; see, e.g., \cite{Greenwood:2008zg}.)
The backbone of the Firewall argument is that at the quantum level, it is impossible to have the entanglement between the black hole interior degrees of freedom and those of the late Hawking radiation - which is necessary for a smooth horizon \cite{Bousso:2012as}.

A more careful examination of the Firewall setup was given in \cite{Hutchinson:2013kka}.
What was shown there is that the near-horizon can never be a vacuum due to the entanglement among {\em all} of the system components, the early and late Hawking radiations and remaining black hole. The quantum effects lead to a non-smooth horizon: the quantum principle of entanglement is fundamentally incompatible  
 with the Equivalence Principle {(more precisely, the interpretation of the Equivalence Principle as implying a smooth horizon)}. The entanglement among {\em all} of the system components will be important in our discussion of BHI in section 4.

BHI can be (and has been) tackled from different angles. For example, in the relatively recent works of \cite{Saini:2015dea} and \cite{Das:2019rex} (that we became aware of sometime after finishing the present work), BHI was studied by employing the functional Schrodinger formalism - which we believe to be the most orthodox framework to study some of the BHI- and Firewall- related issues. The authors analyzed trace of the density matrix squared, and it was shown that once the quantum-field theoretic interaction and correlation are taken into account, unlike in the Hawking's original works, the result was consistent with the unitarity.

\section{Quantum-gravitational framework}

Setting up the full quantum-gravitational framework itself does not require intensive effort: it can be done within the realm of the standard quantum field theory. The framework is essentially scattering amplitude analysis including non-perturbative contributions. As we will see, the non-perturbative contributions are crucial for tackling the black hole information paradox.  Since the perturbative part of the calculation can be handled in the  recently proposed Feynman diagrammatic manner \cite{Park:2019lkh}, we will focus mostly on the non-perturbative contributions.

The boundary dynamics are important both for quantization and BHI \cite{Hatefi:2012bp,Park:2018xtt,Park:2013vpa}. In section 3.1, we qualitatively review the recent result obtained in \cite{Park:2018xtt} in which a reduction called the dimensional reduction to a hypersurface of foliation was carried out as a part of gauge-fixing.\footnote{The gauge-fixing procedure and accompanying projection may have something to do with the notion called the ``nebulon" in \cite{Jacobson:2019gnm}.} This sets the stage for section 3.2 where we outline the scattering amplitude analysis. We write down the scattering amplitude in a schematic notation. A given scattering amplitude factorizes into the perturbative and non-perturbative parts. Although technically complicated, the perturbative part can, in principle, be evaluated. As we will see, the non-perturbative part is reminiscent of the vacuum decay through the Coleman-De Luccia's bounce solution \cite{Coleman:1980aw}\cite{Weinberg}. In section 3.3, we explicitly construct the bounce solution that is based on a shock-wave solution of an Einstein-Hilbert action with an appropriate stress-energy tensor. (Some details can be found in the Appendix.) We discuss the subtleties in regarding our solution as a Coleman-De Luccia's bounce solution. We also discuss the relation of our solution to the ``Lorentzian bounce solutions" of \cite{Ambrus:2005nm,Haggard:2014rza,Christodoulou:2016vny,Bianchi:2018mml,BenAchour:2020mgu,BenAchour:2020gon}. The bounce solution will be of critical subsequent use.

\subsection{Roles of the reduced action as the boundary theory}

Before getting to the quantum-gravitational scattering setup, we review some of the recent developments that we will use in the following subsections and section 4. The main thing to be reviewed is the relevance of the boundary action obtained while quantizing the bulk theory \cite{Park:2019lkh}. In \cite{Park:2019lkh} (and the references therein) it has been noticed that
the boundary theory is naturally obtained in quantizing the bulk theory as a part of gauge-fixing procedure. One can show that as a result of the gauge-fixing, the physical degrees of freedom of the bulk theory are realized on a  hypersurface at the boundary.

 As a matter of fact, the boundary degrees of freedom are a part of the bulk degrees of freedom and the boundary theory is a part of the bulk theory (in that it describes the onshell fluctuations of the bulk geometry; see below). This may sound incompatible with the standard AdS/CFT in which the CFT is a priori independent of the bulk theory. It is not so. Instead, what has been noticed in \cite{Park:2019lkh} (and earlier sequels) should be taken as a refined view of how AdS/CFT works: it must be the way in which the holographic dualities are played out. In the case of the best established AdS$_5$/CFT$_4$ case, it has been shown that the boundary theory, CFT$_4$, can be obtained as the theory of the bulk moduli field \cite{Hatefi:2012bp}.  
Once one considers a given background of the bulk theory, the reduced action describes the fluctuation modes of the physical degrees of freedom. In other words, the boundary theory comes to describe the fluctuations - which is part of the bulk theory physics - around the onshell background geometry. This is why the boundary description is dual to the bulk description: the former {\em is} the latter for the onshell physics. The given boundary configuration determines the boundary conditions of the bulk theory.\footnote{In light of this, it is evident that the standard Dirichlet boundary condition of the bulk gravity theory is one of many possible boundary conditions. As anticipated in \cite{Park:2018xtt} (see \cite{Freidel:2016bxd} for an earlier related observation), one should consider different boundary condition sectors altogether in the enlarged Hilbert space. Or alternatively, consider an inclusive - hopefully most general - boundary condition, if that's technically possible. (Also see the discussion on the quantum boundary modes later.) \la{dbdc}}

The boundary theory plays several significant roles in the discussions that follow: in section 3.2, it is involved in choosing the in- and out- states in the scattering setup. In particular, a given configuration of the boundary theory will serve as the boundary condition of the bulk bounce-type solution that dominates the non-perturbative contribution to the amplitude. In section 3.3 the boundary terms and conditions will be important for vacuum transition probability. In section 4 where we ponder various pieces of information, the boundary theory will be carefully taken into account when scrutinizing the information pertinent to the entanglement between the bulk and boundary components.

\subsection{Scattering setup}

In the quantum gravitational setup, the amplitude to be computed can be written as
\bea
<out;\{\b\}|in;\{\a\}>
\eea
where $|in;\{\a\}>$ and $|out;\{\b\}>$ denote in- and out- states, respectively; $\{\a\},\{\b\}$ collectively stand for various quantum numbers - such as angular momentum, spin, etc - carried by the Fock oscillators. To evaluate the amplitude, one can first compute the corresponding Green's function by following the standard quantum-field theoretic techniques. 
After the Wick contractions, the resulting expression will factorize into   
\bea
\mbox{(Wick contraction part)}<out | in> \la{mstran}
\eea
where $<out | in>$ denotes the vacuum transition part, as we will detail shortly. The in- and out- vacuua, $| in>$ and $ |out>$, will be specified below. While looking deceptively simple, full evaluation of the above amplitude would take considerable efforts. For example, the evaluation of the Wick contraction part involves computing the propagators in a curved background. (But see \cite{Park:2018vci} for use of a flat spacetime propagator in a curved spacetime.) Furthermore, computing the contractions at the loop-levels will necessitate the whole machinery of renormalization procedure. Since these are technical complications that can be implemented in principle, we will focus on the non-perturbative part, $<out | in>$, in the next subsection. 

Let us first examine eq. \rf{mstran} for the different types of information contained. There are three different types.\footnote{Depending the choice of the in-state, there can be information associated with the QCD and electroweak sectors of Standard Model physics. Although such physics will constitute a big chunk of the system information and can be analyzed, it will only introduce unessential complications. To avoid such complications we will consider a simple shock-wave incoming from the asymptotic region.} Firstly, the ``perturbative information" associated with the quantum numbers $\{\a\},\{\b\}$ is preserved (say, through various Kronecker deltas or delta functions) in the usual way of calculating the Green function and S-matrix. Secondly, there is the information coming from quantum corrections. As we will review in section 4, this is the information stored by the quantum boundary modes. These are the modes responsible for a trans-Planckian energy near the horizon \cite{Nurmagambetov:2018het,Nurmagambetov:2019mih}. The third type of information is that associated with the entanglement between the constituents of the black hole and the rest of the system. The preservation of this portion of the information is most subtle to see.

Let us ponder the possible choices for $|in;\{\a\}>$ and $|out;\{\b\}>$. Since we are interested in a quantum-gravitational black hole formation, it is obvious what state to take as final: a certain black hole state. ``Certain" because its hair will depend on the initial state of the boundary. A typical out-state will be the one obtained by applying Fock creation operators on the out-vacuum. 
 For the in-state, one has a large amount of freedom even for infalling matter with spherical symmetry. For simplicity, one may consider choosing a spherical shock-wave infalling from the asymptotic region.

 At this point, it is intriguing to note the relevance of the vacuum decay analysis by Coleman et. al., \cite{Coleman:1977py} and \cite{Coleman:1980aw}, where it was shown that the vacuum decay is mediated by a non-perturbative solution, the so-called  ``bounce." (See, e.g., \cite{Weinberg} for a review.) The vacuum decay amplitude is the probability amplitude for an initial false vacuum to decay into the true vacuum. It is tempting to view the black hole formation process from an initial state as analogous, with the black hole state being the true vacuum {and the white hole state the false vacuum}. One conceptual obstacle to this is the fact that the black hole formation process is not a priori a potential-tunneling process. There are, however, some indications that support the analogy. For instance, it was shown in \cite{Coleman:1980aw} that the vacuum decay of a non-gravitational scalar theory can be embedded in a gravitational setup. 
More importantly, the path integral evaluation of the amplitude above seems to naturally suggest the relevance of a gravitational bounce-type solution. Based on these we propose that the vacuum decay physics under present consideration be viewed as a generalized vacuum transition in the black hole context. (More on this in sections 3.3 and 4.) 

Although there should be diverse initial states that one can take, one choice stands out for its {\em technical} simplicity. (A more realistic solution would be one based on an Oppenheimer-Snyder solution; see, e.g., \cite{BenAchour:2020mgu} and \cite{BenAchour:2020gon}.) Recall that a bounce solution has a bouncing behavior and is time-reversal-symmetric before and after the bounce. A boundary configuration (and thus the corresponding bulk theory boundary condition) that leads to a simple bounce solution is the one dictated by a white hole final state after the bounce. Construction of a bounce solution that corresponds to a more realistic collapsing matter will be much more involved. Fortunately, the essential physics can be understood by taking our bounce solution to which we now turn.

\subsection{A bounce solution}

\begin{figure}
	\centerline{
		\begin{minipage}[b]{10cm}
			\epsfxsize=12cm
			\epsfbox{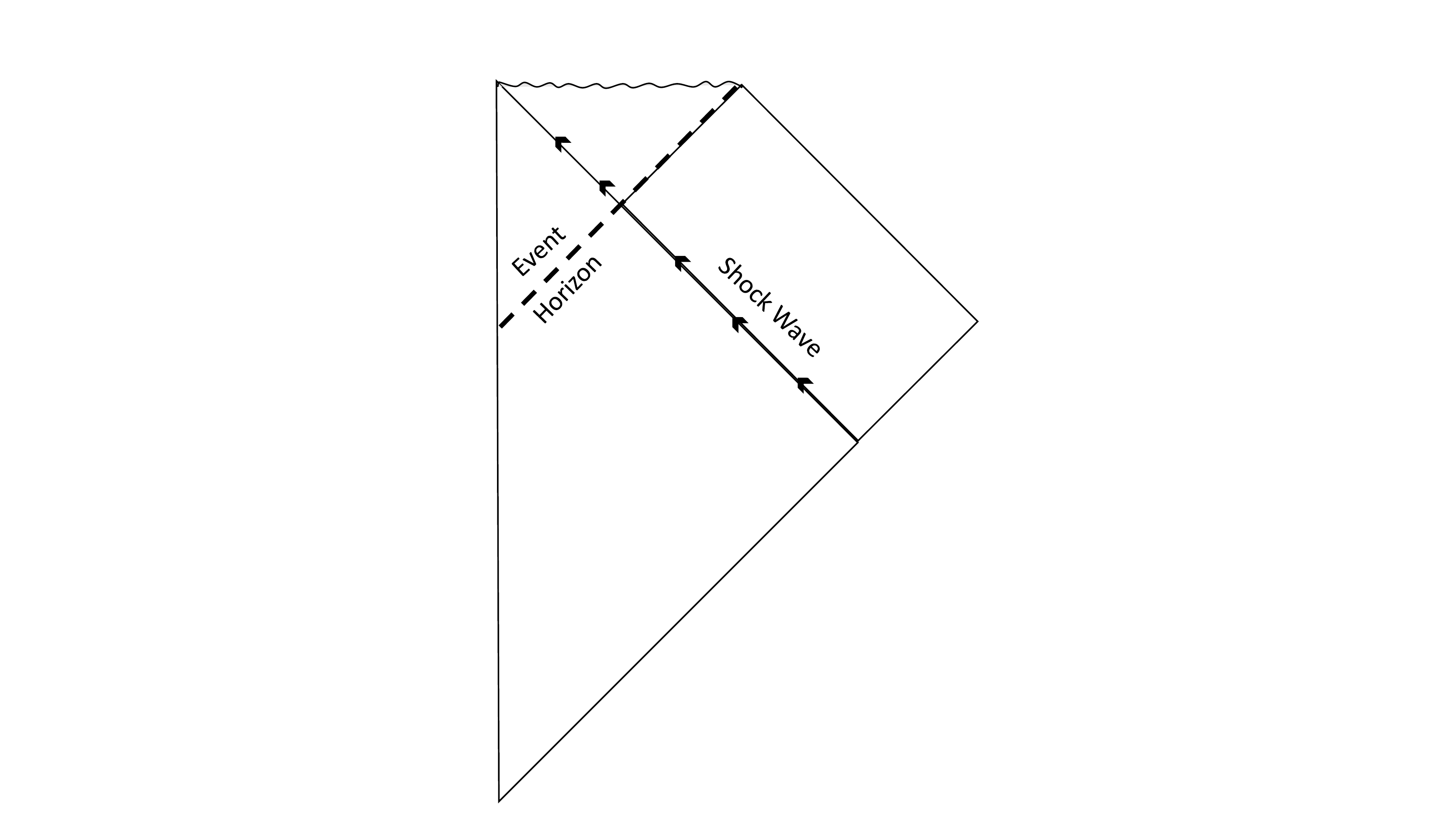}
		\end{minipage}
	}
	\caption{A gravitational collapse of a shock-wave}
	\label{fig}
\end{figure}

Let us construct a gravitational bounce-type solution of a system consisting of an Einstein-Hilbert action and an appropriately defined (i.e., metric-engineered) stress-energy tensor. The solution that we are about to construct is a time-reversal-symmetric version (in the sense of \rf{fsol} below) of a well-known collapse solution whose conformal diagram is given in Fig. 1. As well known \cite{Strominger:1994tn}\cite{Blau}, the following metric and stress-energy corresponding to a collapsing shell (see Fig. 1) satisfies the Einstein equation:
\bea
ds^2 &=& -f(v,r)dv^2+2dudv +r^2 d \Omega^2 \nn\\
T_{vv}&=& \fr1{4\pi G} \fr{m}{r^2} \d(v-v_0)  \la{cs}
\eea 
where $v=t+r_*$ with $r_*$ denoting the tortoise coordinate and $v_0$ denoting a constant. The function $f$ is given by
\be
f(v,r)=1-\fr{2m}r \Th(v-v_0).
\ee
\begin{figure}
	\centerline{
		\begin{minipage}[b]{10cm}
			\epsfxsize=12cm
			\epsfbox{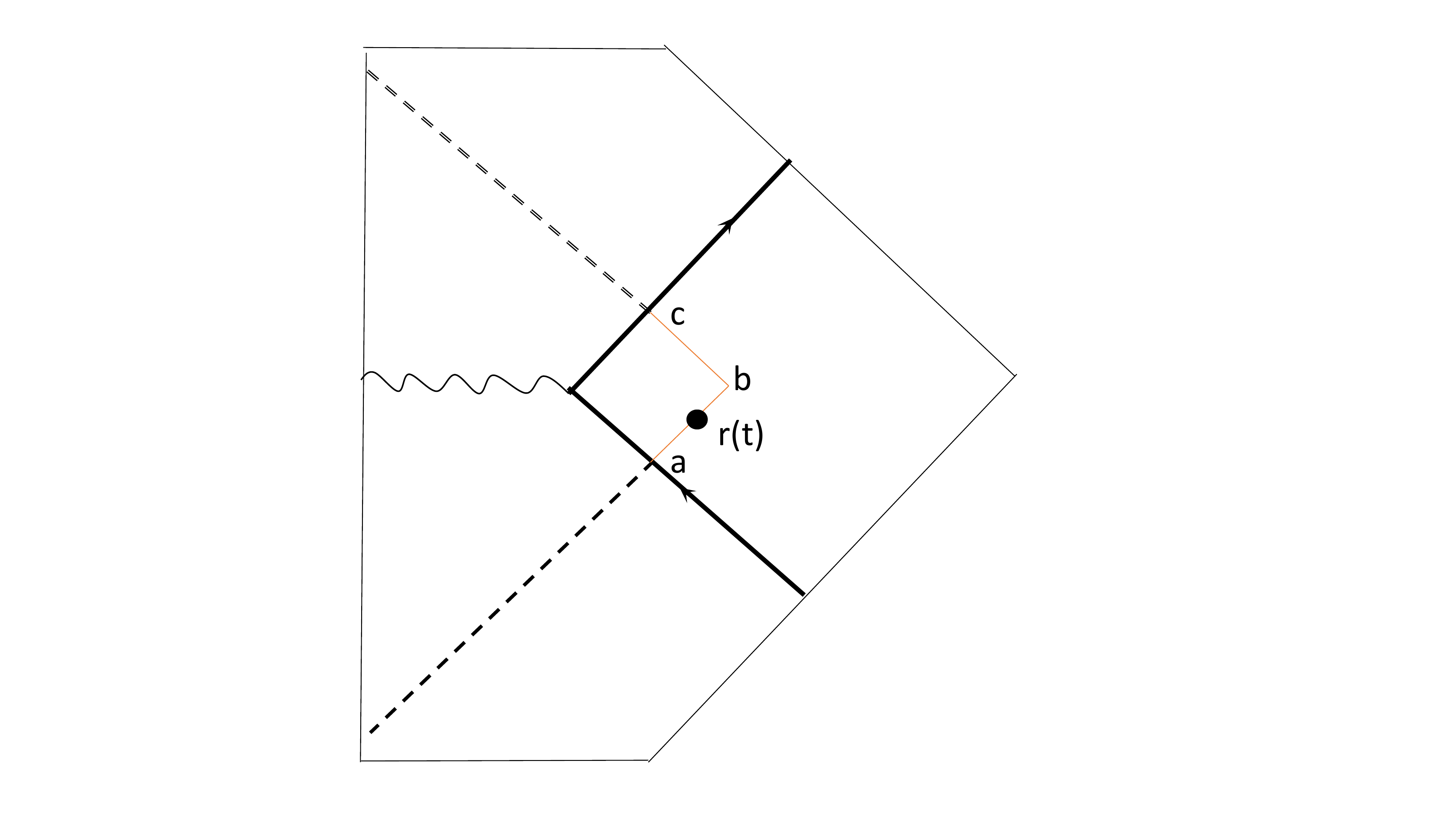}
		\end{minipage}
	}
	\caption{A bounce solution representing a shell that collapses from an initial boundary condition associated with the white hole and subsequently expands. The dotted line represents the event horizon; the segment a-b-c lies outside the shell. The entanglement entropy increases over the a-b segment whereas it decreases over the b-c segment.}
	\label{fig}
\end{figure}
The Heavyside step function $\Th$ takes the values of $0,1$. The bounce solution that we are interested in can be constructed by taking an absolute value of the Schwarzschild time coordinate, $t$: the following metric (see the conformal diagram in Fig. 2)
\bea
ds^2 &=& -F(u,r)du^2  - 2drdu +r^2 d\Omega^2 \la{Fmet}
\eea 
\be
F(u,r)=1-\fr{2m}r \Th(|r_*+u|-r_*-\s_0) \la{fsol}
\ee
where $u=t-r_*$ and $\s_0$ is a positive constant satisfies the metric field equation with the stress-energy tensor given by
\bea
T_{uu}&=& \fr1{4\pi G} \fr{m}{r^2}   |r_*+u|' \delta(r_*-| u+r_*| +\sigma_0) \la{Tuu}
\eea
where the prime $'$ denotes differentiation with respect to the argument. See the Appendix for an explicit check and generalization of the solution. This time-symmetric version of the collapsing shell \rf{cs} will be the dominating non-perturbative contribution to the scattering amplitude.

A word of clarification is in order. The $|in;{\a}>$ state represents a false vacuum state, and this is what was meant by the statement in the introduction that the bonce solution represents a shell that collapses from {\em an initial boundary condition associated with the white hole}. This state is chosen to be the time-reversal of the usual white hole state represented by the upper half of the bounce solution in Fig. 2. (This structure is of course part of how the Coleman's bounce solution \cite{Coleman:1977py} works.)

\subsubsection*{Issues surrounding the bounce solution}

Before analyzing the Wald entropy of the bounce solution \rf{Fmet} in the next section, we examine issues surrounding the bounce solution. Let us start with conceptual ones. To describe the solution we are using the term ``instanton-type" contribution in the sense that the bounce solution makes a non-perturbative contribution. However, its Euclidean origin is not entirely clear. One may consider an Euclidean counter-part by analytically continuing the time coordinate $t$ with resulting $u,v$. For its full justification, however, it will be necessary to consider matter part, such as a scalar field, as we further comment on below. Another conceptual aspect is that since we are considering the bounce as a non-perturbative contribution to the path integral, it is a classical solution. One can of course try to obtain the quantum-corrected bounce solution by solving the 1PI action once the action is obtained (in a truncated form). Presumably this approach may well qualitatively confirm the results obtained along the line of \cite{Ambrus:2005nm} and \cite{Haggard:2014rza} where solutions that reflect quantum effects were constructed. 

One technical issue is the undesirable feature of our bounce solution: containing $|t|$, the solution is not analytic at $t=0$. Although it should be possible, at least as an ad hoc measure, to realize the bounce solution as a limit of a sequence of, e.g., Sigmoid functions, and avoid the feature, one may also try to avoid such a singularity by constructing a less artificial and more realistic solution. As a matter of fact, this appears to be what has been achieved in \cite{BenAchour:2020mgu} and \cite{BenAchour:2020gon} by constructing a bounce solution based on an Oppenheimer-Snyder solution.\footnote{We also expect that it will be possible to construct a bounce solution of a more realistic system, such as an Einstein-scalar system, that does not share such a behavior.} As discussed therein, Planckian scale physics is expected near the transition point.  

Lastly, since the bounce solution mediates the vacuum decay, one can analyze the probability of the transition. As a matter of fact, this issue has been addressed to some extent, e.g., in \cite{Christodoulou:2016vny,Bianchi:2018mml}. In particular, in \cite{Bianchi:2018mml} it was noted on dimensional grounds that the tunneling probability $p$ should be given by
\be
p\sim e^{-\fr{m^2}{m_{Pl}^2}} \la{prop}
\ee
where $m_{Pl}$ denotes the Planck mass. However, the evaporation time that goes with this probability - which is an inverse of \rf{prop} - is far smaller than the Page result, $\sim m^3$, or the result obtained in \cite{Haggard:2014rza}, $\sim m^2$, based on the so-called `classicality parameter.' In fact, the evaporation time based on \rf{prop} grossly misses the boundary conditions and dynamics: the result \rf{prop} is appropriate only for special set of boundary conditions.

To set the stage to see this, we first demonstrate how \rf{prop} may be realized in the present setup. A rigorous verification of \rf{prop} will require {\em systematic} analysis by employing an Euclidean bounce as well as finite temperature setup, which will not be pursued here. It is nevertheless possible to sketch the steps that lead to verification of \rf{prop}. Let us first recall that the gravitational action can be written as (see, e.g., \cite{Poisson})
\be
S= S_{EH}+S_{YGH} +S_0 \la{tac}
\ee
where 
\bea
S_{EH}&=&\fr1{16\pi}\int_{\cal V}d^4x \sqrt{|g|}\; R\nn\\
S_{YGH}&=&\ve\fr1{8\pi}\int_{\pa{\cal V}}d^3y \sqrt{|h|}\; K \nn\\
S_0&=&\ve\fr1{8\pi}\int_{\pa{\cal V}}d^3y \sqrt{|h|}\; K_0. 
\eea
$S_{YGH}$ is the York-Gibbons-Hawking term (YGH) and $S_0$ denotes $S_{YGH}$ evaluated for a flat background; $\ve$ is $\ve=1$ for timelike ${\pa{\cal V}}$ and $\ve=-1$ for spacelike ${\pa{\cal V}}$. Compared to the usual practice, it is necessary choose $\ve=-1$, since an Euclidean action is being considered ultimately. Before evaluating the action in \rf{tac}, a cautionary remark is in order. The role of the YGH term is to impose a Dirichlet boundary condition. There is, however, a subtle issue \cite{Park:2018xtt} of whether or not the bounce solution (for that matter, the shock-wave solution \rf{cs}) actually satisfies the Dirichlet boundary condition: it would seem more natural for it to satisfy a certain Neumann-type boundary condition. Setting this issue aside for now, we substitute $g_{\m\n}=bounce$ into $S_{YGH}$ to evaluate the YGH term for the bounce solution. By going to the ADM formalism and the standard manipulations, one gets, for the Euclidean action, 
\be
S=S_{bulk}- \fr{2}{16\pi G}\int d\t \int d^2\th N \sqrt{\s}\,(k_b-k_0)
\ee
where $N$ denotes the lapse function and $k_0,k_b$ are the extrinsic curvatures associated with the two-spheres embedded in the respective spaces, flat and bounce. The action can be evaluated over the bounce solution. The bulk term vanishes and the nonzero contribution comes from the boundary term: by taking the 2D surface at an asymptotic region outside the shell, one gets 
\be
- \fr2{16\pi G}\int d\t \int d^2\th N \sqrt{\s}\,(k_b-k_0) \sim \fr1{G} \int d\t\, Gm.
\ee
It is reasonable to expect that the relevant temperature is Hawking temperature: $\int d\t \sim Gm$. Considering the overall minus sign in front of the Euclidean action, one gets 
\be
 \sim -Gm^2 \la{Dygh}
\ee
thus confirming \rf{prop}. In the remainder, we argue that the analysis above is incomplete for the reason that the corresponding analysis must be carried for a whole tower of different boundary conditions.  

The analysis above is based on the Dirichlet boundary condition, and as previously mentioned, one must be careful with the boundary conditions. The importance of boundary conditions has been noted in several different contexts. For instance, it was noted in \cite{Chen:1998aw} that there are potentially infinitely many different conditions with associated boundary terms, and quasi-local charges take different values depending on them. As mentioned in the introduction and footnote \ref{dbdc}, for quantization the Hilbert space must be expanded to include all of the different sectors of possible boundary conditions. One of the implications of the recent works \cite{Park:2017wiw,Park:2016fxc,Park:2018xtt} is that the well-known Dirichlet boundary condition is very special in that is should be associated with the ``vacuum" of the 3D dynamics.
Having a common bulk solution, all of the different boundary conditions must be associated with different aspects of the bulk fluctuation reflected on the boundary \cite{Park:2018xtt}. This status of the matter suggests that in calculating a physical quantity, such as the decay time, one must take an ``ensemble" average over the boundary conditions. Also, in light of the analysis in \cite{Park:2018xtt} in which it was shown that in general Noether charges are not conserved for non-Dirichlet boundary conditions, the mass will not be conserved for a generic boundary condition. Since the mass value will depend on the boundary terms and conditions, it will not, in general, turn out to be the same as the mass parameter $m$ for a non-Dirichlet type condition. Instead it is natural to expect the value to be smaller, reflecting the mass loss by Hawking radiation. It will be of great interest to see whether the ``ensemble" average may lead to the $m^2$- or $m^3$- scaling of the decay time.

\section{Black hole hair}

With the discussions in section 2 and 3, we are ready to take an in-depth look at the system information and its preservation. Examination of the information should be carried out at various stages of the quantum-level analysis. In particular, the information stored in the boundary configuration becomes relevant at the early stage of quantization when determining the physical states, as well as later when choosing the in- and out- states. The next stage is the actual evaluation of the scattering amplitudes. The perturbative piece of information - which is associated with the Fock space creation operators - can be extracted in the standard manner - though highly complicated technically. 

As for the non-perturbative information, there are two types. The first non-perturbative information is associated with the instanton-like solution called the ``bounce." The name was originally introduced in a non-gravitational quantum field theory, in fact, in the quantum mechanical context \cite{Coleman:1977py}. In the gravitational context, the same name was given to a class of solutions that have a ``bouncing" feature \cite{Ambrus:2005nm}\cite{Haggard:2014rza}. It appears that the connection between the two types of the bounce solutions was not recognized in the literature. We propose that the gravitational bounce solution be viewed as having the same origin as the original bounce solution: it is associated with the vacuum transition.     
The second type of the nonperturbative information can be extracted by analyzing the 1PI effective action. Except for the fact that the object to be analyzed is the 1PI action, the analysis is in the same spirit as in quasi-normal mode calculation: the analysis employs the black hole perturbation method. What information can be extracted through such analysis? For this, it is useful to start with the classical-level analysis. The presence of nontrivial matter field configurations represents part of the system information.  In addition, the standard classical black hole perturbation reveals that the radiation takes away all of the multipoles except the first several ones associated with mass, charge, and angular momentum. With these approaches elevated to the quantum-level, one should discover that the quantum boundary modes carry a significant amount of the system information (see below). This should be the quantum-gravitational bleaching anticipated in \cite{Park:2013rm}.

Below we scrutinize each of these two sectors. In particular, for the information mediated by the bounce solution, we examine the entanglement  between the black hole degrees of freedom and the rest of the system. The entanglementinformation should be measured by the Wald's entropy charge.     
It is important to note that in considering the bounce solution - which is time-dependent - one can nevertheless consider the timelike Killing vector {\em after} the shell crosses with the expanding event horizon. This is because according to Birkhoff theorem, the geometry outside the expanding spherical region should still be the static Schwarzschild geometry. A shifted Page-like curve is generically produced. Elevating the Wald's entropy charge to the quantum-level is straightforward: it is the vev of the charge which can be calculated by applying the Wald's method to the quantum-corrected action.

\vspace{.1in}

In section 4.1, the information mediated by the bounce solution is analyzed. In section 4.2, the recent results in black hole perturbation are reviewed with a focus on the information carried by the boundary quantum modes. In section 4.3, we combine the results and present an overall account of black hole formation and its information release pattern.

\subsection{Information mediated by a bounce}

We have discussed that the dominant contribution to the non-perturbative part of the scattering amplitudes will come from a bounce solution. For a convenient visual illustration with minimum technical complications, one can use the null-shell-collapse-based bounce solution explicitly constructed in the previous section. Ideally, one should consider an Einstein-Hilbert action coupled with matter, say, a scalar field, instead of the stress-energy tensor engineered by the choice of the metric. Then the in- and out- states would belong to different non-perturbative sectors of the theory. What we propose is that the transition between two states belonging to different sectors may be analyzed along the line of \cite{Coleman:1977py}. In fact,  in the case of the field theory analysis in \cite{Coleman:1980aw}, it was shown that the vacuum decay physics of a non-gravitational scalar theory - which is a transition from a false vacuum of the potential to the true vacuum - can be embedded in a gravitational setup. In any event, it should be possible to construct a bounce solution in an Einstein-scalar system. (For instance, a collapsing-shell solution of an Einstein-scalar system was numerically constructed in \cite{Nunez:1998gj}.) It is just that we employ the bounce solution constructed in the previous section to avoid inessential technical complexities.

Recall that as we have reviewed in section 3.1, the boundary condition of the bulk theory is tied with the boundary theory state and that boundary theory describes the physical degrees of freedom of the bulk theory in the given background. As we saw in section 3, a technically simple choice of the boundary state is associated with the bounce solution that has the black hole state as the true vacuum and the white hole state as the false vacuum. In the path integral evaluation of the amplitude, one should consider expanding the action around the bounce solution. The leading-order action, i.e, the action evaluated on the bounce solution, can be pulled out of the path integral; it is the leading non-perturbative contribution.

Another role of the bounce solution is that being a time-dependent geometry, it generates particles \cite{Rubakov:1984pa}\cite{Kraus:1994zu}. It is not entirely surprising since such particle generation is well known in cosmology \cite{Parker:1968mv}\cite{Birrell}. Moreover, it is those particles that were identified as the (now-known-as) Hawking radiation in the Hawking's original work. That interpretation should remain valid in the present non-perturbative context: the particles generated by the vacuum decay should be the Hawking radition.\footnote{In \cite{Kraus:1994zu} it was shown that the presence of the bounce solution the Hawking radiation is essentially unaffected. In the present work, the particle generation caused by the bounce {\em is identified} with the Hawking radiation.}

Toward the end of this section, we will carry out entropy calculation and observe a shifted Page-like curve originating. Let us try to develop some intuitions before getting to the quantitative analysis. When considering the entanglement among the system components, it is crucial to take into account the modes represented by the Fock oscillators since they too will be entangled with the Hawking radiation. 
 The entanglement between the black hole degrees of freedom and the rest of the system - including not only the Hawking modes {\em but also the hard (and soft) modes} - must be considered. This is because everything in the initial infall must have been entangled, and thus should remain as such. (See \cite{Park:2014hya}\cite{Semenoff:2019dqe}\cite{Tomaras:2019sjq} for recent related works.) As stated previously, this must be the reason for the fact that the resulting curve is shifted as compared to the Page curve.

Lastly, being too simple, our bounce solution misses some aspects of the physics associated with black hole evaporation. A more realistic solution as well as to properly take the quantum gravitational corrections into account. (The fact that one considers the instanton-type contribution through the bounce solution means that non-perturbative corrections are being taken into account. In addition to this, one must consider $\hbar$-corrections to the bounce solution itself.) To capture such physics it will be crucial to start with a more realistic solution that we describe in the next subsection.

\subsection{Nonperturbative hair revealed by BH perturbation}

Another line of the system information can be extracted through the standard black hole perturbation theory. At the classical level, black hole perturbation theory was developed long ago. (See, e.g., \cite{Frolov,Maggiore} for reviews.) The classical quasi-normal mode (QNM) analysis reveals that a black hole radiates away most of its multipole characteristics, except the ones associated with mass, charge, and angular momentum. It should be possible to extend the black hole perturbation theory to the quantum-level. The repetition of the QNM analysis should again show that the higher-order multipoles will be radiated away.

By changing the methodology somewhat, one can explore different aspects of the system information. The method is quite suited when it is the quantum-corrections that induce the perturbations. Applied to the present case, one may try to obtain the quantum-corrected bounce solution by solving the quantum-corrected action. However, although the bounce solution serves well the purpose of analyzing the first type of the non-perturbative contribution, there are simpler solutions for illustrating the second-type non-perturbative contribution presently being considered: they are the familiar static or stationary solutions such as a Schwarzschild or Kerr solution. Thus we leave the bounce solution in the remainder of the present subsection.

Let us review the recent results obtained by deforming a familiar stationary solution by time-dependent perturbations.    
In \cite{Nurmagambetov:2019mih}, the following action of an Einstein-Maxwell-scalar system was considered:
\bea
&&\hspace{-.2in}S=\fr1{\k^2}\int d^4x \sqrt{-g}\;\Big[R-2\L\Big] +\int d^4 x \sqrt{-g}\;  \Big[c_1  R^2+c_2 R_{\m\n}R^{\m\n} +\cdots\Big]  \nn\\
&&\hspace{-.3in}-\fr14 \int d^4x \sqrt{-g}\;F_{\m\n}F^{\m\n}  -\int d^4 x \sqrt{-g}\;\Big[|\pa_\mu \psi-iqA_\m \psi|^2 
+{\l}\Big(|\psi|^2+\fr{1}{2\l} \n^2\Big)^2   \Big].
  \la{emsactcasetwo}
 \nn\\
\eea
A time-dependent solution that represents a deformation of a Kerr black hole was constructed by taking the following ansatz:
\bea
ds^2&=&-\fr{F(t,z,\th)}{z^2}(dt+a \sin^2\th d\f)^2+2(dt+a \sin^2\th d\f)\Big(-\fr{dz}{z^2}+a\sin^2\th d\f\Big)  \nn\\
  && +\Phi^2(t,z,\th) (d\th^2+\sin^2d\f^2)  \nn\\
&=&-\fr{F(t,z,\theta)}{z^2}dt^2-\fr2{z^2}dtdz+2a\Big(-  \fr{F(t,z,\theta)}{z^2}+1\Big)\sin^2\th\, dtd\f -\fr{2a}{z^2}\sin^2\th dzd\f \nn\\
&&+\Phi^2(t,z,\theta) d\th^2+\Big( -\fr{a^2F(t,z,\theta)}{z^2}\sin^2\th+2a^2\sin^2\th+\Phi^2(t,z,\theta) \Big)\sin^2\th d\f^2. 
\la{Kerrans}
\nn\\
\eea
with 
\bea
\hspace{-.5in}F(t,z,\th)&=& F_0(t,\th) +F_1(t,\th) z+ F_2(t,\th)z^2+F_3(t,\th)z^3 + ...\nonumber\\
&+&\k^2  \Big[F_0^h(t,\th) +F_1^h(t,\th) z+ F_2^h(t,\th)z^2+F_3^h(t,\th)z^3 + ...\Big], \nn\\
\Phi(t,z,\th)&=&\frac{1}{z}+\Phi_0(t,\th) +\Phi_1(t,\th) z+ \Phi_2(t,\th)z^2+\Phi_3(t,\th)z^3 + ...\nonumber\\
&+& \k^2  \Big[\fr{\Phi_{-1}^h(t,\th)}{z}+\Phi_0^h(t,\th) +\Phi_1^h(t,\th) z+ \Phi_2^h(t,\th)z^2+\Phi_3^h(t,\th)z^3 + ...\Big] \nn\\
   \la{1stans}
\eea
where the modes with superscript `$h$' represent the quantum modes. There are similar expressions for the matter fields. The subsequent analysis showed that, while the perturbation settles down and thus has time-dependence, one gets a trans-Planckian energy near the horizon. Although the framework is not suitable to exhibit the multipole characteristics being radiated away, such information-carrying radiation should, in analogy with the QNM analysis, accompany the ring-down procedure.  
In the series expansion above, it is straightforward to impose a quite general, if not the most general, boundary condition: the presence of the dynamic boundary modes, such as $\Phi_{-1}^h(t,\th),\Phi_0^h(t,\th))$, implies that the solution satisfies a certain non-Dirichlet boundary condition. Consideration of such a solution brings two returns. Firstly, it brings out the role of the boundary quantum modes. Secondly and not unrelatedly, it leads to a trans-Planckian energy near the event horizon.

The motivation of \cite{Nurmagambetov:2018het,Nurmagambetov:2019mih} was the Firewall. The tripartite system that split out of the original black hole makes it clear that one cannot have the the required entanglement for a smooth horizon between the late Hawking radiation and remaining black hole degrees of freedom \cite{Hutchinson:2013kka}. An infalling observer will observe a highly blue-shifted radiation near the horizon if the area is not vacuum. The results in \cite{Nurmagambetov:2018het,Nurmagambetov:2019mih} seem to be a computational confirmation of the presence of a Firewall-type structure at the horizon.

\subsection{Information-release pattern}

In the previous sections, we have collected several ingredients: in section 3 we have defined system information in order to study the evolution (namely, the formation and evaporation) of a black hole in a quantum-gravitational scattering setup. We also have explicitly contructed a bounce solution that determines the non-perturbative contribution in the path integral. In subsections 4.1 and 4.2 we have proposed that the particle generation under the vacuum decay be identified as the Hawking radiation and part of the system information is captured by various black hole perturbation analyses in the spirit of the QNM modes. One gets the following picture of black hole evolution as an upshot of putting these ingredients together. Initially, an incoming shell with perturbative excitations by the Fock creation operators starts collapsing, and at the same time the perturbative excitations - whose information is obviously preserved - are scattered away. While the collapsing body forms a black hole, most of the multipole moments except the ones associated with the mass, charge, and angular momentum are radiated away. Therefore, much of the system information heads for the asymptotic region. Once the black hole is formed, it starts to decay through the instanton-type non-perturbative channel, the bounce. In this subsection we examine, as the last piece of the complete picture, the information associated with the entanglement between the system components.

Although the information preservation seems obvious in the present picture, one should understand why the semi-classical analysis had led to the apparent information loss. It is essentially because in the semi-classical analysis only the black hole and Hawking radiation (early and late) were considered. As we have seen, when considering the entanglement among the system components it is important to consider {\em all} of the components. In other words, for a proper understanding of the purity and unitary evolution of the system, the pre-Hawking radiation and boundary dynamics must be taken into account as well \cite{Park:2013rm,Park:2017wiw}. There is an indication of this from another angle in the entropy analysis below.

The presence of the pre-Hawking radiation suggests that it should play a role in the entanglement among the system components. We will come back to this below but for now let us see how a (shifted) Page-like curve - which was presented in \cite{Page:1993wv} as the hallmark of preservation of the information associated with the entanglements - is produced by using the Wald's entropy \cite{Wald:1993nt}. Let us review the fact that in a gravitational theory, there exists an offshell current $J^\n(\xi^\m)$ (whose explicit expression can be found, e.g., in \cite{Padmanabhan:2009vy}) that is associated with the gauge parameter $\xi^\m$. If the background geometry admits a timelike Killing vector, the current will lead to a familiar conserved charge when integrated over a spatial volume. Interestingly, the current $J^\n$ admits an expression through a two-form $J^{\r\s}$: $J^\m=\nabla_\n J^{\m\n}$, so the Noether charge $Q$ is
\bea
Q=\int d^3\Sigma_\m J^\m=\fr12 \int d^2\Sigma_{\m\n}J^{\m\n}
\eea 
with
\bea
J^{\m\n}=\fr1{16\pi} (\nabla^\m \xi^\n-\nabla^\n \xi^\m).
\eea
The entropy, defined as $S=\fr{2\pi}{\k}Q$ (the factor $\fr{2\pi}{\k}$ is inserted to make $S$ consistent with the Bekenstein-Hawking entropy), takes 
\bea
S=\fr{A_H}{4}
\eea
where $A_H$ denotes the area of the horizon.
Although the bounce solution is a time-dependent geometry, one can nevertheless use the same formula for the reason explained in the introduction of this section with the understanding that, for the bounce solution, the surface $\Sigma_{\m\n}$ is time-dependent: it increases until the event horizon reaches the final size and decreases afterwards. (One may wonder how a Noether charge could be time-dependent. As discussed in \cite{Park:2018xtt} (see section 3.2), this has to do with the boundary conditions.) During the time interval in which the shell remains outside of the event horizon, the segment a-b-c in Fig. 2, the entropy formula above, once applied to $\Sigma_{\m\n}$, generically leads to a Page-like curve.\footnote{While this manuscript was in preparation, \cite{Almheiri:2019qdq} and \cite{Penington:2019kki}, in which the relevance a worm hole solution for the Page curve was recognized, appeared. The recent works in which the Page curve has been obtained include \cite{Penington:2019npb,Almheiri:2019psf,Almheiri:2019yqk,Rozali:2019day,Chen:2019uhq,Bousso:2019ykv}.} The are $A_H$ is given by
\be
A_H=4\pi r^2(t)
\ee
The entanglement entropy increases over the a-b segment whereas it decreases over the b-c segment, the hallmark characteristic of the Page curve. (It is also evident from this that a generic bounce solution will display this feature.) This can be seen at an analytical level as follows. The time-dependent radial coordinate $r(t)$ is determined once $r_*(t)$ is. In the b-c segment,
$v=t+r_*=v_0>0$\footnote{The a-b-c segment is in the fourth quadrant in the $(v,u)$ coordinate system, hence $v_0>0$ for b-c segment. Similarly, one has $u_0<0$ for a-b segment.} is fixed, and thus:
\bea
r_*=-t+v_0
\eea 
In this region $r_*$ decreases - so does $r$ - as $t$ increases. In the a-b segment, $u=t-r_*\equiv u_0<0$ is fixed and one gets
\bea
r_*=t-u_0
\eea
Thus $r_*$ and $r$ increase as $t$ increases.

The quantum extension of the Wald entropy formula seems straightforward: instead of the classical action, one should use the quantum-corrected action.\footnote{One subtlety is that the Birkhoff theorem may not be applicable at the quantum level. Then the geometry will be time-dependent in general and there will be no timelike Killing vector. More work is required but what saves the situation should be that existence of an asymptotically timelike Killing vector must be sufficient. Namely, it should be possible to compute the entropy by integrating over the surface at the boundary.} Since the quantum-corrected solution is expected to preserve the bouncing feature of the classical solution, so should the quantum-corrected Page-like curve.

One significant implication of the analysis above is the relevance of the pre-Hawking radiation \cite{Park:2013rm} - which is governed by the boundary theory. The analysis above shows that the black hole formation is a process in which the to-be-black hole degrees of freedom and the rest of the system are redistributed\footnote{{Of course, redistribution includes particle creation processes.}} while remaining entangled and thus keeping the initial purity. As the shell collapses the event horizon appears and its area increases: this means that the black hole degrees of freedom becomes more and more entangled with the rest of the system while the whole system remains pure. In the bouncing-out phase of the shell, as more and more Hawking radiation (namely, the particles created by the decay) joins the bulk, the bulk degrees of freedom almost constitute the overall state - which is pure - with less entanglement with the black hole degrees of freedom.   
This is reflected in the decrease of the area of the horizon.\footnote{The entropy changes associated with the time-dependence of the bounce solution seem consistent with the teleological nature of a black hole. It was anticipated in \cite{Park:2014mba} that the teleological property of the event horizon of a black hole will be crucial in BHI.}

Although the present picture of black hole formation and evaporation shares qualitative features with the Page's account, it refines it in that the entropy starts to increase once the event horizon appears and to decrease once it reaches its full size. If we disregard the entanglement between the pre-Hawking modes and black hole mode, the entropy should increase for a while even after the black hole formation. However, if we take the pre-Hawking modes into account, there are already substantial degrees of freedom out, so the Hawking radiation makes the entanglement decrease.

\section{Conclusion}

In this work we have systematized black hole evolution in a quantum-gravitational scattering setup in which all of the different types of information are kept track of throughout. The results obtained can be summarized into the following three: the first is the qualitative picture (presented in the first paragraph of section 4.3) of the system evolution. The second is the construction of the explicit bounce solution and its role in BHI. The third is the computation in regards to what the Wald entropy measures and its quantum extension. In the present quantum-gravitational scattering approach, it is evident that the information is preserved. The entanglement between the black hole degrees of freedom and the rest of the system can be measured by the Wald's entropy charge; a shifted Page-like curve has been produced.

Compared with the semi-classical analysis, the present study has several key ingredients that have led to the information preservation. The first ingredient is the role of the boundary degrees of freedom. In the recent works, it has been anticipated that consideration of non-Dirichlet boundary conditions will be crucial for resolving BHI. It was proposed that the Hilbert space should be enlarged to include the sectors of different boundary conditions. The in- and out- states that we have considered belong to the different sectors, and this naturally led to the consideration of a bounce solution. The second ingredient is in conjunction with the bounce solution that we have come up with: the generalization of the vacuum transition. For this, the general configuration space should be considered. Also crucial was the identification of the particles generated from the vacuum transition with the Hawking radiation. 
On a more technical side, employing the simple collapsing-shell-based bounce solution has facilitated the analysis. 
We believe that the bounce solution that we have constructed, though not entirely realistic as it may be, captures the essence of the problem.\footnote{{It is noteworthy (and ironic) that the black hole formation process contains a seed of its destruction through the bounce-type non-perturbative contribution.}} As for a more realistic bounce solution, we take the works of \cite{Coleman:1980aw} and \cite{Nunez:1998gj} as an indication that the stress-energy tensor found in the present work can be realized in the context of an Einstein-scalar system.

One of the lessons learned is that the non-perturbative quantum effects (the instanton effect in the present case) are important regardless of large curvature. The fact that the quantum effects are important regardless of large curvature was also well demonstrated in the recent works where a trans-Planckian energy results near the horizon of the time-dependent black hole \cite{Nurmagambetov:2018het,Nurmagambetov:2019mih}.

\vspace{.3in}
There are several future directions:
\vspace{.1in}

One intriguing aspect about the Noether current is the fact that certain components of the current are radially conserved \cite{Liu:2017kml}. Presumably there exists a flow of entropy whose flux is conserved \cite{Freedman:2016zud}\cite{Bakhmatov:2017ihw}\cite{Oh:2017pkr}. It will be interesting to understand potential implications of this for the Ryu-Takayanagi's result \cite{Ryu:2006ef}.   

As we have noticed in the body, there are certain similarities between the Hawking modes and QNMs. Two things hamper a direction identification of these two notions. Firstly, the Hawking radiation is mainly a q-number description whereas QNMs are in the context of a c-number description. Secondly, the QNMs do not constitute a complete set since they satisfy special boundary conditions \cite{Ching:1995tj} whereas the Hawking modes provide a basis. 
In spite of these differences, the quasi-normal modes should somehow be related to the Hawking modes \cite{Corda:2012tz}, and making the relation more precise will be of some interest. 

Another direction concerns construction of a more realistic bounce solution. The bounce solution that we have considered is time-reverse-symmetric. There may be other types of non-perturbative solutions that are not. Perhaps not unrelated, it will be interesting to construct a quantum-corrected bounce and see whether or not the solution will remain time-reverse-symmetric. 
As discussed in section 3.3, the transition probability has room for significant improvement. The task is tied with enumeration of all possible boundary conditions, which is an interesting problem on its own.

Our final thoughts are on the following rather radical possibility. In the main body we have seen that the non-perturbative part of the scattering amplitude is dominated by the bounce solution. The bounce solution is obviously a time-dependent solution. Although we have not demonstrated it for a bounce-type solution, for a geometry obtained by perturbing a stationary geometry, such as a perturbed Kerr, we have recently demonstrated (as reviewed in the body) that generic perturbations - which should thus be time-dependent - lead to a trans-Planckian energy. Perhaps a Firewall may be a way for Nature to avoid the curvature singularity. In other words, Nature may avoid the curvature singularity by quantum effects that produce a Firewall so that the singularity cannot be accessed.

\newpage
\appendix

\renewcommand{\theequation}{A.\arabic{equation}}
\setcounter{equation}{0}

\section{Some details of bounce solution}

Another way of writing the solution \rf{fsol} 
\be
F(u,r)=1-\fr{2m}r \Th(|r_*+u|-r_*-\s_0) \la{fsolq}
\ee
is
\bea
 F = \left\{ \begin{array}{ll}
1-\fr{2m}r \Th(u-\s_0) & \mbox{if $t \geq 0$};\\
1-\fr{2m}r \Th(-v-\s_0) & \mbox{if $t < 0$}.\end{array} \right. 
\la{fsola}
\eea
This notation makes it clearer that the solution is expressed in terms of $(v,r)$ for $t<0$.
Since $t\geq 0$ ($t<0$) solution has the same form of the collapsing shell (expanding shell), eq. \rf{fsola} should satisfy the field equation. (Nevertheless it is explicitly checked below.) Ultimately, one would be interested in a more realistic solution. In particular, it will be of interest to see whether the solution above can be generalized by replacing the $\Th$ function by something more realistic. For that, it is useful to examine the Einstein tensor associated with the metric \rf{Fmet}, first without using the explicit form of F. We illustrate it for $t \geq 0$ region:
\bea
G_{\m\n}=\left(
\begin{array}{cccc}
	-\frac{\left(r \pa_r F-1\right) F+r \pa_u F+F^2}{r^2} & \frac{r \pa_r F+F-1}{r^2} & 0 & 0 \\
	\frac{r \pa_r F+F-1}{r^2} & 0 & 0 & 0 \\
	0 & 0 & \frac{1}{2}  \pa_r (r^2 \pa_r F) & 0 \\
	0 & 0 & 0 &   \sin ^2\theta \;\frac{1}{2} \pa_r (r^2 \pa_r F) \\
\end{array}
\right). \nn\\
\eea
This can be rewritten as
\bea
G_{\m\n}=\left(
\begin{array}{cccc}
-F G_{ur}-\frac{ \pa_u F}{r} & G_{ur} & 0 & 0 \\
	G_{ur} & 0 & 0 & 0 \\
	0 & 0 & G_{\th\th} & 0 \\
	0 & 0 & 0 &   \sin^2\th\; G_{\th\th} \\
\end{array}
\right) \la{seten} \nn\\
\eea
with
\bea
G_{ur} = \frac{ \pa_r(r F)-1}{r^2},\quad G_{\th\th} = \frac{1}{2} \pa_r (r^2 \pa_r F) 
\la{Gur}
\eea
As mentioned in the beginning, the form of F in \rf{fsola} makes it obvious that it is a solution. Let us still examine each entry in \rf{seten} and see whether that provides some clues for a more general solution that may preserve some of the nice features of \rf{fsola}. Consider $G_{ur}$:
\bea
G_{ur} &=& \frac{ [\pa_r(r -\Th)]-1}{r^2}=-\frac{ \pa_r\Th}{r^2}.
\eea
This vanishes since $\Th=\Th(u)$ or $\Th=\Th(v)$: 
\bea
G_{ur} &=& 0.
\eea
Similarly one can show that $G_{\th\th}$ vanishes:
\bea
G_{\th\th}=\frac{1}{2}  \pa_r (r^2 \pa_r F) =m  \,\pa_r \Th=0.
\eea 
$G_{\f\f}$ is proportional to $G_{\th\th}$, and thus it is not necessary to consider $G_{\f\f}$.
Finally, note that with $G_{ur}=0$, $G_{uu}$ simplifies to
\bea
G_{uu} =-\frac{ \pa_u F}{r}
\eea
which is then canceled, in the field equation, by $T_{uu}$ given in \rf{Tuu}.

The steps above does seem to provide a useful guidance when one tries to construct a more general form of F than our toy model. For instance, the examination above naturally suggests the following form: 
\bea
F = \left\{ \begin{array}{ll}
	1-\fr{K_1(u)}{r} & \mbox{if $t \geq 0$};\\
	1-\fr{K_2(v)}{r}  & \mbox{if $t < 0$}.\end{array} \right. 
\la{fsolgs}
\eea
Then the $(u,r)$ or $(v,r)$ sector of $G_{\m\n}$ will have to be matched with the corresponding block of the stress tensor constructed out of, say, a scalar system. Construction of a realistic solution will be further pursued elsewhere. In such a case one may have to pay a close attention to the boundary terms (see \cite{Esposito} for a general review).

\newpage

\end{document}